\documentclass[10pt,prl,aps,twocolumn,floatfix,showpacs]{revtex4}
\usepackage{graphicx}
\usepackage{amssymb}
\usepackage{bm}

\begin{document}

\title{Evidence for deconfined quantum criticality \\
in a two-dimensional Heisenberg model with four-spin interactions}

\author{Anders W. Sandvik} 
\affiliation{Department of Physics, Boston University, 
590 Commonwealth Avenue, Boston, Massachusetts 02215}

\date{\today}

\begin{abstract}
Using ground-state projector quantum Monte Carlo simulations in the valence bond basis, 
it is demonstrated that non-frustrating four-spin interactions can destroy the N\'eel order 
of the two-dimensional $S=1/2$ Heisenberg antiferromagnet and drive it into a valence-bond 
solid (VBS) phase. Results for spin and dimer correlations are consistent with a single 
continuous transition, and all data exhibit finite-size scaling with a single set of exponents;
$z=1, \nu=0.78 \pm 0.03$, and $\eta=0.26 \pm 0.03$. The unusually large $\eta$ and an 
emergent $U(1)$ symmetry, detected using VBS order parameter histograms, provide strong 
evidence for a deconfined quantum critical point.
\end{abstract}


\maketitle

Since the discovery in 1986 of high-$T_{\rm c}$ superconductivity in layered cuprates, quantum 
phase transitions in two-dimensional (2D) antiferromagnets have formed a central topic in
condensed matter physics \cite{chn,sachdevbook}. While superconductivity is induced in the 
CuO$_{\rm 2}$ planes of the cuprates by doping with charge carriers, other mechanisms for destroying 
the N\'eel order and stabilizing different ground states have also been intensely investigated 
theoretically. Considerable efforts have been devoted to possible spin liquid 
("RVB" \cite{and87}) and valence-bond solid (VBS) states driven by magnetic frustration 
\cite{rea89,j1j2old,mis05}. This work has been partially motivated by the hope that 
an understanding of generic features of quantum phase transitions in 2D antiferromagnets could shed 
light also on the mechanisms at work in the cuprates \cite{sac03}. Quantum fluctuation driven 
phase transitions are also of broader relevance in the context of strongly correlated 
systems \cite{sondhi97}. 

A quantum phase transition occurs as a function of some parameter at temperature $T=0$ and
corresponds to a $T>0$ transition in an effective classical system with an imaginary-time 
dimension---the path integral  \cite{herzandsuzuki}. 
The standard Landau-Ginzburg-Wilson framework for critical phenomena should thus be applicable, with the 
dimensionality $d \to d+z$, where the dynamic exponent $z$ depends on the way space and time correlations are 
related. In the paradigm prevailing until recently, the ``Landau rules" for the nature of the 
transition---continuous or first-order---were also assumed to remain valid for quantum phase transitions. 
A direct transition between two ordered phases should thus be generically first-order if two 
different symmetries are broken. This notion has recently been challenged by Senthil {\it et al.}, who 
argued that quantum phase transitions separating two ordered phases can be generically continuous, even 
when different symmetries are broken \cite{sen04}. This theory of ``deconfined" quantum critical 
points was first developed for the transition between an antiferromagnetic (AF) and a 
valence-bond-solid (VBS) state. Both these states have confined $S=1$ excitations---gapless 
magnons and gapped "triplons", respectively. The critical point is 
characterized by deconfined $S=1/2$ spinons coupled to an emergent $U(1)$ gauge field \cite{sen04}. 
In 2D the deconfined state is unstable and exists only at a point separating the two ordered phases. 
The AF and VBS order parameters arise as a consequence of spinon confinement. In this Letter, 
quantum Monte Carlo (QMC) results are presented which support this theory. 

Preceding the theory of deconfined quantum critical points, continuous transitions between two 
ordered quantum states had been suggested based on numerical simulations \cite{assaad,san02}. 
However, in more detailed studies following the theoretical developments it has 
proved difficult to confirm their existence. Instead, many studies have pointed to weakly 
first-order AF--VBS transitions \cite{kuk0405,kuk06,kagome,mel06,sirker} or other  
scenarios inconsistent with deconfined quantum criticality \cite{san06}. To date, large-scale QMC studies 
of potential deconfined quantum critical points have focused on spin (or hard-core bosonic) models with 
spin-anisotropic interactions \cite{kuk0405,kuk06,kagome,mel06}. Frustrated $SU(2)$ (Heisenberg) 
symmetric interactions, which cannot be studied using QMC simulations due to the infamous
"sign problem", have been considered in exact diagonalization studies \cite{poilblanc}. Because
of the limitations to very small lattices, it has not been possible to study phase transitions 
in detail, however. In fact, not even the nature of the VBS state has been completely settled in 
basic models such as the J$_{\rm 1}$-J$_{\rm 2}$ frustrated Heisenberg model \cite{j1j2new}. 

Here it will be shown that the AF order of the square-lattice Heisenberg model can be destroyed 
also by non-frustrated isotropic interactions accessible to QMC simulations. The following 
Hamiltonian will be discussed:
\begin{equation}
H = J\sum_{\langle ij\rangle} \mathbf{S}_i \cdot \mathbf{S}_j -
Q\sum_{\langle ijkl\rangle} (\mathbf{S}_i \cdot \mathbf{S}_j-\hbox{$\frac{1}{4}$})
(\mathbf{S}_k \cdot \mathbf{S}_l-\hbox{$\frac{1}{4}$}),
\label{jqham}
\end{equation}
where $\langle ij\rangle$ denotes nearest-neighbor sites and $\langle ijkl\rangle$ refers 
to the corners of a plaquette, such that $ij$ and $kl$ form two parallel adjacent horizontal or 
vertical links. This interaction contains a subset of the four-site ring-exchange, and with $Q>0$ 
there is no QMC sign problem. Note that the purpose here is not to model any specific material, but
simply to construct a model system in which an AF--VBS transition can be investigated. It will be 
shown below that the ground state of the J-Q model has AF order for $J/Q \agt 0.04$ and VBS order 
for $J/Q \alt 0.04$.

\begin{figure}
\includegraphics[width=6.15cm, clip]{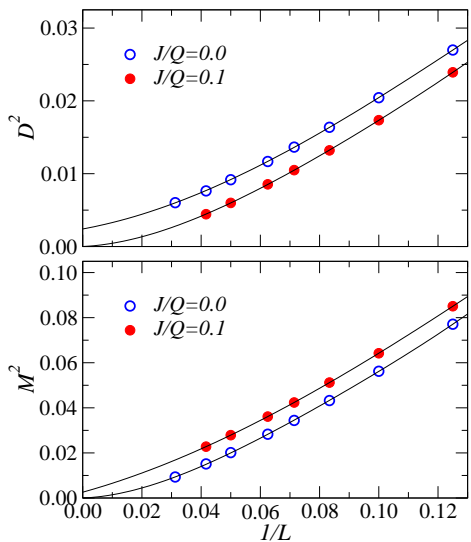}
\caption{(Color online) Finite size scaling of the squared spin ($M$) and dimer ($D$)
order parameters at $J/Q=0$ and $0.1$. The curves are cubic fits. Statistical errors are much 
smaller than the symbols.}
\label{fig1}
\vskip-5mm
\end{figure}

To study the ground state of the hamiltonian (\ref{jqham}), an approximation-free projector technique 
in the valence bond basis \cite{vbmc} is employed which is ideally suited for multi-spin interactions 
formed out of singlet projection operators $(\mathbf{S}_i \cdot \mathbf{S}_j-\hbox{$\frac{1}{4}$})$.
Here $L\times L$ lattices with $L$ up to $32$ are considered. Larger systems may be reachable 
using loop algorithms in the standard $S^z$ basis, which have been used for $U(1)$ models with 
four-site interactions \cite{jkmethod,errornote}. The valence bond basis has its advantages, 
however, including an improved estimator for the singlet--triplet gap.

Results will be presented for spin-spin ($s$) and dimer-dimer ($d$) correlation functions,
\begin{eqnarray}
C_s(\mathbf{r}) & = & \langle \mathbf{S}(\mathbf{0}) \cdot \mathbf{S}(\mathbf{r})\rangle, 
\label{cs} \\
C_d(\mathbf{r}) & = & \langle [\mathbf{S}(\mathbf{0}) \cdot \mathbf{S}(\hat \mathbf{x})]
[\mathbf{S}(\mathbf{r}) \cdot \mathbf{S}(\mathbf{r} + \hat \mathbf{x})]\rangle,
\label{cd}
\end{eqnarray}
where $\hat \mathbf{x}$ denotes a lattice unit vector in the $x$ direction. The AF order parameter
is the staggered magnetization, the square of which is calculated;
\begin{equation}
M^2 = \frac{1}{N}\sum_{\bf r} C_s(\textbf{r})(-1)^{r_x+r_y}.
\end{equation}
The VBS state can have either columnar or plaquette order, both of which break $Z_4$ symmetry. An 
important aspect of the theory is that these order parameters should both exhibit divergent fluctuations 
at the deconfined critical point. Only at some length-scale diverging as a power 
of the correlation length should one of them be singled out \cite{sen04}. This is analogous to the 
irrelevance of $Z_4$ anisotropy in the 3D XY model \cite{jose} and corresponds directly to the 
predicted emergent $U(1)$ symmetry. The $\mathbf{q}=(\pi,0)$ 
dimer order parameter,
\begin{equation}
D^2 = \frac{1}{N}\sum_{\bf r} C_d(\textbf{r})(-1)^{r_x},
\end{equation}
is divergent for both columnar and plaquette VBS order and will be studied here.

Extrapolations of the AF and VBS order parameters, shown in Fig.~\ref{fig1}, demonstrate that 
there is long-range VBS order but no AF order  at maximal four-spin interaction; $J/Q=0$ (note that there 
are still two-site interactions present when $J=0$; simulations for $J < 0$ are 
sign problematic). Also shown are results at $J/Q=0.1$, where the situation is the reverse;
there is AF order but the VBS order vanishes. Thus there is an AF--VBS transition
somewhere in the range $0 < J/Q < 0.1$, or there could be a region of AF/VBS coexistence (which would 
be analogous to a supersolid state). The nature of the VBS order---columnar or plaquette---is not
clear from these calculations. However, simulations of open-boundary rectangular lattices, in which 
a unique columnar or plaquette pattern can be stabilized \cite{san02}, indicate that columnar order is preferred. 
The extrapolated VBS correlation at $J/Q=0$ is $D^2 \approx 0.0024$.

\begin{figure}
\null\includegraphics[width=6.05cm, clip]{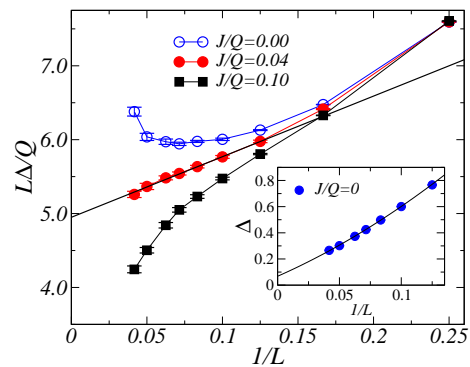}
\caption{(Color online) Finite size scaling of the singlet-triplet excitation gap multiplied by 
the system size $L$. The behavior at $J/Q=0.04$ corresponds to $z=1$.}
\label{fig2}
\vskip-5mm
\end{figure}

The deconfined theory has dynamic exponent $z=1$ \cite{sen04}. 
This exponent can be directly accessed through the finite-size scaling of the singlet--triplet gap; 
$\Delta \sim L^{-z}$. To demonstrate consistency with $z=1$, the scaling of $L\Delta$ is shown in 
Fig.~\ref{fig2} for $J/Q=0$ and $0.1$, as well as for $J/Q=0.04$ which will be shown below to be close 
to criticality. Here $L\Delta$ extrapolates to a non-zero value, supporting $z=1$, and at $J/Q=0$ and 
$0.1$ the behaviors are what would be expected off criticality. The inset of Fig.~\ref{fig2} shows an 
infinite-size extrapolation of the gap at $J/Q=0$, giving $\Delta/Q\approx 0.07$.

\begin{figure}
\includegraphics[width=6.25cm, clip]{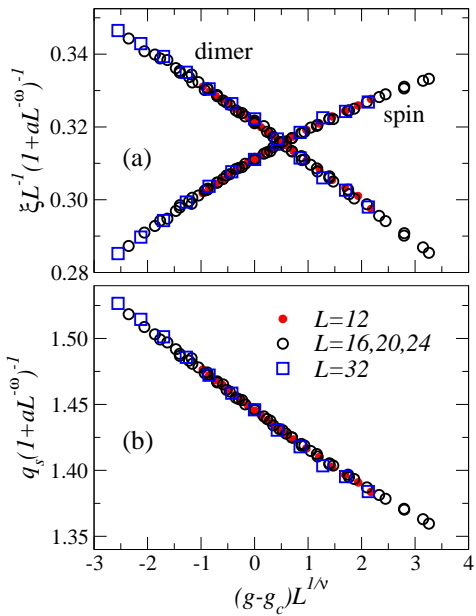}
\caption{(Color online) Scaling with $\nu=0.78$ and $g_c=0.04$ of the correlation lengths (a) 
and the spin Binder ratio (b).} 
\label{fig3}
\vskip-5mm
\end{figure}

Correlation lengths $\xi_s$ and $\xi_d$ for spins and dimers are defined in the standard way as the 
square-roots of the second moments of the correlation functions (\ref{cs}) and (\ref{cd}). 
Also useful is the Binder cumulant, defined for the 
spin as $q_s=\langle M^4\rangle/\langle M^2\rangle^2$. Finite-size scaling of these quantities is used to extract 
the critical coupling and the correlation length exponent $\nu$. To achieve good data collapse, a subleading 
correction is also included. With $g=J/Q$, the scaling ansatz is,
\begin{equation}
A(g,L)=L^{\kappa}(1+aL^{-\omega})f[(g-g_c)L^{1/\nu}],
\label{scaling}
\end{equation}
where $A=\xi_s,\xi_d$, or $q_s$, and $\kappa=1$ for $\xi_s,\xi_d$ and $0$ for $q_s$. As seen 
in Fig.~\ref{fig3}, these quantities can be scaled using $g_c=0.040 \pm 0.003$ and a common $\nu=0.78\pm 0.03$. 
In all cases, the subleading exponent $\omega \approx 2$, and the scaling is nearly as good if 
$\omega=2$ is fixed throughout. Interestingly, the best prefactor $a$ is then almost equal for 
$\xi_s$ and $\xi_d$, $a\approx 8$, but this may be coincidental.

\begin{figure}
\null~~\includegraphics[width=6.cm, clip]{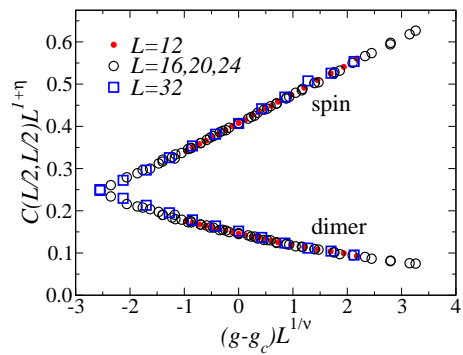}
\caption{Long-distance spin and dimer correlations scaled using $\nu=0.78$, $\eta=0.26$, 
and $g_c=0.04$.}
\label{fig4}
\vskip-5mm
\end{figure}

Next, the correlation functions $C_{s,d}(\mathbf{r})$ at the longest lattice distance, $\mathbf{r}=(L/2,L/2)$,
are analyzed to extract the correlation function exponent $\eta$. The expected scaling is as in 
Eq.~(\ref{scaling}) with $\kappa=-(1+\eta)$. Now $g_c$ and $\nu$ are kept fixed at the values determined 
above. As shown in Fig.~\ref{fig4}, a single exponent describes both the spin and dimer data, and in this 
case a subleading correction is not needed ($a=0$). The exponent, $\eta = 0.26 \pm 0.03$, is unusually 
large. In the 3D $O(3)$ universality class, describing transitions between the AF and a featureless 
gapped state \cite{chn,lingwang}, $\eta\approx 0.04$. A larger $\eta$ for deconfined quantum criticality 
was argued for on physical grounds by Senthil {\it et al.} \cite{sen04}. The universality class was argued
to be that of the hedgehog suppressed $O(3)$ transition, for which $\beta/\nu=(1+\eta)/2=0.80\pm 0.05$ 
was obtained in simulations of a classical model in \cite{mot04}. This is larger than $\beta/\nu=0.63 \pm 0.02$ 
found here, but on the other hand smaller lattices were used in \cite{mot04} and there may also be issues with 
how hedgehogs were suppressed. The direct study of an actual AF--VBS transition presented above can thus be 
expected to be more reliable.

\begin{figure}
\null~~~\includegraphics[width=5.5cm, clip]{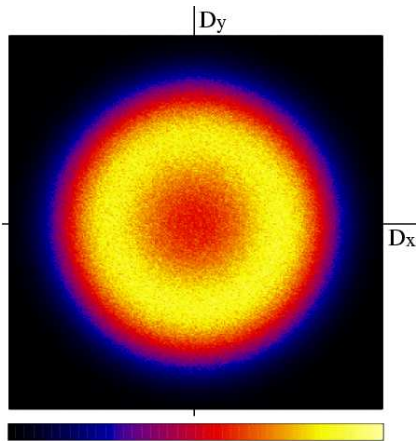}
\caption{(Color online) Histogram of the dimer order parameter for an $L=32$ system at $J/Q=0$. 
The ring shape demonstrates an emergent $U(1)$ symmetry, i.e., irrelevance of the $Z_4$ 
anisotropy of the VBS order parameter.}
\label{fig5}
\vskip-5mm
\end{figure}

It is also interesting to study the probability distribution $P(D_x,D_y)$ of the dimer order 
parameter. In the VBS phase, one would expect this distribution to reflect the $Z_4$ symmetry, 
i.e., for a columnar VBS there should be peaks at $D_x=0,D_y=\pm D$ and $D_x=\pm D,D_y=0$
(whereas a plaquette state would give rise to peaks rotated by $45^\circ$). It should be noted, however, 
that $P(D_x,D_y)$ is a basis dependent function. In the valence bond simulations \cite{vbmc} the 
order parameters used to construct $P(D_x,D_y)$ are matrix elements
(with $\hat \mathbf{e} =  \hat \mathbf{x}, \hat \mathbf{y}$),
\begin{equation}
D_{e} = \frac{\langle \Psi_b |
\frac{1}{N}\sum_r \mathbf{S}(\mathbf{r}) \cdot \mathbf{S}(\mathbf{r}+\hat \mathbf{e})(-1)^{r_e}
|\Psi_a\rangle}{\langle \Psi_b |\Psi_a\rangle},
\end{equation}
where $|\Psi_a\rangle$, $|\Psi_b\rangle$ are valence bond states generated by operating with a 
high power $H^n$ on trial state (stochastically sampling valence bond evolutions). 
Although $P(D_x,D_y)$ is not a physically measurable quantity, any symmetry detected in
it should reflect an underlying symmetry of the projected state. Fig.~\ref{fig5} shows 
a color-coded $P(D_x,D_y)$ histogram generated at $J/Q=0$. The expected $Z_4$ symmetry 
of the VBS is not seen; instead the histogram is ring shaped, which indicates a $U(1)$ 
symmetric order parameter. Such an emergent $U(1)$ symmetry is in fact predicted \cite{sen04} by the 
deconfined theory in the VBS phase below a length scale $\Lambda$ which diverges faster than the VBS 
correlation length; $\Lambda \sim \xi_d^a$, with $a>1$. Thus, inside the VBS phase, if the system length 
$L \ll \Lambda$ one should expect to find an $U(1)$ symmetric order parameter, with the $Z_4$ becoming 
relevant only for larger sizes (and then seen as four peaks emerging in the histogram). Here, apparently, 
even at $J/Q=0$ the system is close enough to the critical point for the system length ($L=32$) to
be less than $\Lambda$ and, hence, $Z_4$ to be irrelevant. Recalling that the VBS gap is small, 
$\Delta/Q \approx 0.07$, and that $\Lambda \sim \xi_d^{a} \sim \Delta^{-a}$, this seems
reasonable. On moving closer to the critical point, $P(D_x,D_y)$ smoothly evolves into a single 
broad peak centered at $(0,0)$, as is expected for a continuous transition. Note that the finite-size 
extrapolation of the order parameter in Fig.~\ref{fig1} is not sensitive to the nature of the VBS 
state---plaquette or columnar---and should give the correct magnitude of the order parameter even 
though no $Z_4$ features are yet seen in the histogram for these system sizes. 

The above analysis points consistently to a deconfined quantum critical point as the most
likely scenario for the AF--VBS transition in the J-Q model. One set of exponents describes
both spin and dimer correlations, the value of $\eta$ is unusually large, and there is an emergent 
$U(1)$ symmetry in the VBS order parameter. In principle one cannot rule out a weakly 
first-order transition on the basis of finite-size data. However, although the lattice sizes used 
in this work are not extremely large, it would be hard to explain why a first-order transition should 
lead to the kind of scaling observed. A narrow region of AF/VBS coexistence is also unlikely, as there 
would then be two transitions and there is no reason to expect the spin and dimer critical exponents 
to be the same (in particular, the unusually large $\eta$). It is difficult to say anything more 
quantitative regarding a possible first-order transition or coexistence based on the calculations 
presented here.

An emergent $U(1)$ symmetry may also explain why it has been so difficult to determine the nature 
of the VBS state in the J$_{\rm 1}$-J$_{\rm 2}$ Heisenberg model \cite{j1j2new}. Even if the transition 
would be weakly first-order in this case \cite{sirker,kruger}, an emergent $U(1)$ symmetry could still 
affect small lattices, thus making it difficult to distinguish between columnar and plaquette VBS
patterns. Emergent $U(1)$ symmetry may be more common than deconfined quantum criticality and could 
hence affect many models with VBS states. The high density of low-lying singlets associated with 
$U(1)$ symmetry may also affect exact diagonalization studies \cite{mis05} of level spectra.

{\it Acknowledgments.---}
I would like to thank I. Affleck, L. Balents, K. Beach, M. P. A. Fisher, K. Harada, N. Kawashima, R. Melko, 
O. Motrunich, N. Prokof'ev, S. Sachdev, D. Scalapino, T. Senthil, B. Svistunov, and A. Vishwanath for 
stimulating discussions. This work was supported by the NSF under grant No.~DMR-0513930.

\null


\begin{thebibliography}{00}
\bibitem{errornote} 
A claim in \cite{vbmc} of sign problems in the $S^z$ basis is wrong.
\end{thebibliography}
\end{document}